# Spatio-Temporal Photonic Metalattice

Wallace Jaffray[1†], Sven Stengel[1†], Farhan Ali[2], Mustafa Goksu Ozlu[3],
Colton B. Fruhling[3], Maria Antonietta Vincenti[4,5], Domenico De Ceglia[4,5],
Michael Scalora[4,5], Alexandra Boltasseva[3], Vladimir M. Shalaev[3],
Andrea Di Falco[2], Marcello Ferrera[1*]

[1]Institute of Photonics and Quantum Sciences, Heriot-Watt University, Edinburgh, EH14 4AS, UK

[2]School of Physics and Astronomy, University of St. Andrews, St Andrews, UK

[3]Elmore Family School of Electrical Computer Engineering and Birck Nanotechnology Center,
Purdue University, West Lafayette, Indiana, USA

[4]Department of Information Engineering,University of Brescia,Brescia, Italy

[5]CNR - National Institute of Optics, Via Branze 45, 25123 Brescia, Italy

*Corresponding author. Email: m.ferrera@hw.ac.uk

†These authors contributed equally to this work.

**When coherent light interacts with an ordered lattice whose periodicity is comparable to its wavelength, constructive interference produces a diffraction pattern as in crystallography, where x-rays are employed to reveal atomic structures. By asking 'when' the diffractive object exist, rather than 'where', we implicitly introduce time as a design parameter, thus enabling the creation of spatio-temporal metalattices. In these structures, temporal modulation of optical properties complements the spatial patterning, unlocking advanced functionalities such as dynamic reconfigurability, nonreciprocal behavior, coherent amplification, and tailored spectral response. However, for these effects to be relevant an extreme temporal modulation of the refractive index is necessary. In this work, we realize a two-dimensional spatio-temporal metalattice by integrating a physically patterned spatial modulation with an orthogonal temporal lattice induced by interfering ultrafast pulses, using highly nonlinear, low-index transparent conducting films. While the optical pumps experience a uniform medium, the lattice emerges through a strongly enhanced and internally generated third harmonic signal. The transient lattice shows comparable diffraction efficiency to the physical structure and is also dynamically reconfigurable via a broad range of parameters, including pump pulse delay, incidence angle, and wavelength, offering exceptional versatility for ultra-**

**fast transient lithography and photon manipulation in both momentum and frequency. This approach shifts device design from fixed fabrication constraints to radiation engineering, opening new pathways towards ultrafast reconfigurable photonics.**

Transparent conductors, which are widely recognized in industry for their electro-optic roles in photovoltaic and touch screen technologies (*1*), have recently been investigated for their nonlinear optical responses within spectral regions where their refractive index falls well below unity (*2–4*). In this regime, under ultra-fast optical excitation, an unprecedented temporal refractive index gradient has been observed. This phenomenon underpins several remarkable experimental outcomes, including unitary refractive index modulation (*5*), broadband spectral routing of ultra-fast pulses (*6*), dynamic manipulation of photon energy and momentum (*7–9*), and a generalized form of Snell's law for non-reciprocal metasurfaces where a temporal phase gradient complements the conventional spatial gradient (*10, 11*).

In addition, the unprecedented speed of the material response, capable of tracking few-cycle optical excitations (*12, 13*), has sparked renewed interest in exploring optical phenomena within time-varying systems. Particular attention has been directed towards emerging concepts such as photonic time crystals (*14–17*) and ultra-efficient coherent amplification (*18*). This is well justified by the opportunity to finally overcome the longstanding trade-off between amplitude and speed in nonlinear processes (*19*). With time now effectively incorporated into the engineering toolbox for designing advanced optical materials and metamaterials, a new frontier has emerged: the development of systems capable of ultrafast manipulation of optical radiation across the classic and quantum regimes (*8, 20, 21*). Motivated by this paradigm shift, we turn our attention to diffraction, arguably the most fundamental light-matter interaction, to examine its behavior in the presence of an optically-induced lattice pattern on a low-index material, and to directly compare it with standard scattering from spatially structured media.

In this context, transient diffraction gratings have established themselves as a powerful spectroscopic tool in materials science and chemistry (*22–24*). By interfering two phase-coherent pump beams at oblique angles, a spatial modulation of the refractive index can be inscribed in a film, which subsequently diffracts a probe into different diffraction orders. This technique has enabled detailed studies of electronic and vibrational dynamics in complex materials, as well as molecular processes (*25–27*), and has revealed first-order diffraction in vanadium dioxide ($VO_2$) films under ultraviolet excitation (*28*). More recently, it has also been shown that transient gratings induced in $VO_2$ can drive guided-mode resonances in thin-film metasurfaces, thereby extending their role from spectroscopy to ultrafast, reconfigurable photonic functionality (*29*).

Here we unify two distinct concepts into one single optical system: a vertical ridge grating, patterned onto an aluminum zinc oxide (AZO) film, and a horizontal transient temporal grating of comparable periodicity, formed by the interference of near-infrared (NIR) beams. This spatio-temporal structure remains almost "invisible" to the longer wavelength pump radiation as the grating periodicity is very close to the incident wavelength, yet it is revealed by the internally generated and locally enhanced third harmonic (TH) signal. The recorded diffraction pattern reveals the underlying spatio-temporal lattice responsible for its formation, much like how x-ray diffraction exposes atomic arrangements. By introducing time as a design parameter in the synthesis of metamaterials, we significantly expand the dimensionality of the accessible parameter space, thus unlocking a broader spectrum of functionalities and application domains. In our specific implementation, the spatio-temporal framework can be precisely engineered by tuning the wavelength of the incident pump beams, their angle of incidence, the temporal delay between the pumps or any combination of the these.

Therefore, the novelty of the proposed approach lies along four fundamental lines: i) this work represents an experimental and direct comparison between the scattering properties of a physical structure and those of an optically induced counterpart (*30*); ii) we demonstrate the creation of a temporal grating, generated via ultrafast two-beam interference, in a reversible manner and within a solid state system (*31, 32*); iii) although grating devices can generally provide an improvement of nonlinear optical processes, the proposed system offers a strong broadband TH enhancement which eliminates the need for external probing (*33*); iv) the suggested technology offers an exceptionally versatile platform for ultrafast reconfigurable metasurfaces, leveraging a broad array of control parameters that operate on femtosecond timescales (*34–36*). These mechanisms are readily adaptable to a wide spectrum of advanced applications, including dynamic metasurface design, high resolution spectral analysis, ultra-fast wavelength and spatial division multiplexing, and the on-chip engineering of high-harmonic generation. Moreover, many of these directions point toward potential relevance in the quantum domain, where analogous principles may inform or enable emerging quantum photonic architecture and use-cases (*37*).

## Experimentals results

Experiments were conducted using the setup reported in Fig. 1a. Here, a femtosecond laser source operating at 10 Hz emits tunable ultrafast pulses (1250 nm-1600 nm, 85 fs duration) which are split into equal-intensity beams along separate propagation arms, one incorporating a variable delay stage. The two attained NIR beams are then focused and spatially overlapped on the sample surface. This generates a TH signal with sufficient wavelength resolution to produce a double diffraction pattern in the far field which is due to both the static grating along the y-axis, and the optically induced transient grating along the x-axis. Experiments were conducted at different incident angles $\theta_F$ (6° and 20°) using balanced pump intensities. All beams were p-polarized (i.e., electric field lying in the plane of incidence). Measurements were conducted in both single- and double-beam configurations on a plain 500 nm-thick AZO film and on a focused ion beam–fabricated, 1D fully etched metagrating (metalattice) with a groove density of $\approx$ 560 lines/mm and a duty cycle of 0.7. Figure 1a inset shows a SEM figure of the fabricated sample. The spatial distribution of TH signals for all four configurations is shown in Fig. 1b. More specifically, in panel (1) a single pump beam, incident on the unpatterned film, generates a collinear TH. In panel (2) a single pump incident on the metagrating produces two additional TH beams, diffracted symmetrically above and below the central TH signal. The vertical emission angles of these diffraction orders are consistent with standard grating theory applied to the TH wavelength. Panel (3) shows the result of two beams incident on the unpatterned film, yielding four spatially separated TH signals whose origin will be explained in details in the following paragraphs. Finally, panel (4) displays the full spatio-temporal diffraction pattern produced by two pump beams incident on the metagrating, which leads to fourteen distinct TH spots in the far field, revealing the embedded two-dimensional lattice produced by the combined static and transient structures.

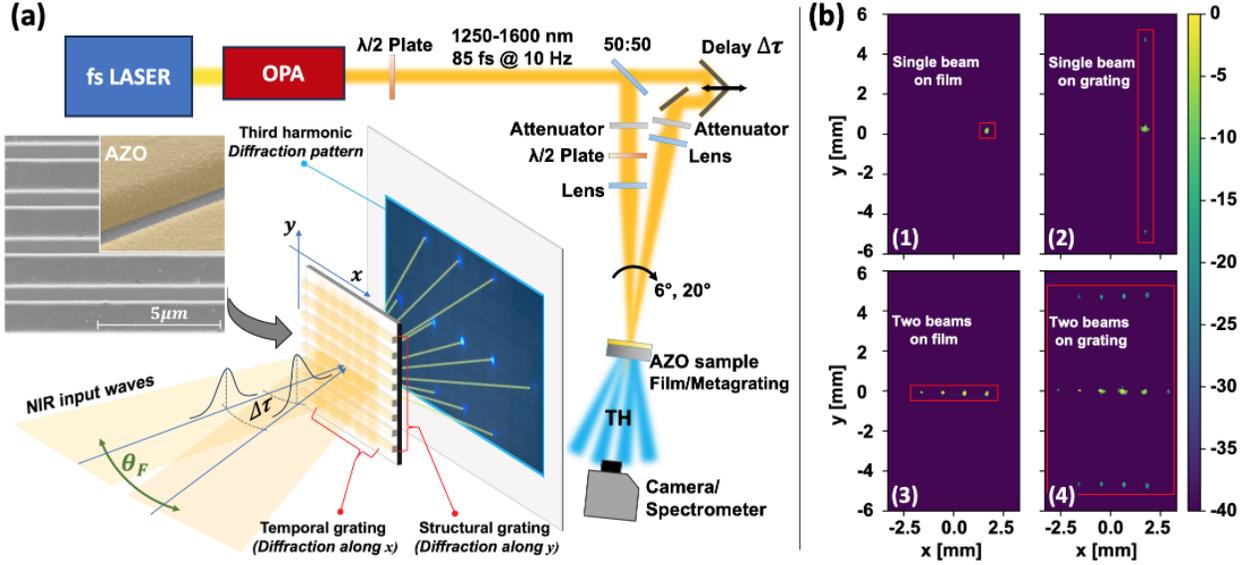

**Figure 1**: **Spatio-Temporal metagrating.** (a) Experimental set-up and magnified schematic of the meta-grating sample, inclusive of an SEM image of the metagrating. Two NIR pump beams interact with a vertically oriented ridge grating, photolithographically patterned onto a thin film of aluminum zinc oxide. The system is excited with 85 fs pulses at NIR wavelengths. Interference between the two pump beams generates a transient horizontal grating, orthogonal to the static vertical structure. (b) Experimentally recorded TH output intensity patterns captured via a visible CCD camera for four excitation configurations: (1) Single beam on unpatterned AZO film; (2) Single beam on meta-grating; (3) two beams on unpatterned film; (4) two beams on meta-grating. Panel (1) shows standard THG from the plain AZO surface. Panels (2) and (3) isolate the diffraction effects arising from the temporal and structural gratings, respectively. Panel (4) reveals the full spatio-spectral diffraction pattern resulting from the combined static and transient lattice.

Special attention must be given to the TH pattern recorded in Fig. 1b-3, where the observed TH signals can be attributed to two fundamental mechanisms: i) photon momentum (wavevector) conservation pertaining to the TH process (*13*); ii) direct diffraction from the transient temporal grating formed by two-beam interference. These two contributions are now examined in detail.

Regarding point i), the THG process involves the mixing of three pump photons, yielding four distinct k-vectors according to the vector diagram in Fig. 2a-1 and Fig. 2a-2, which are aligned with the four TH signals recorded in Fig. 1b-3. Two of these TH k-vectors arise from collinear summations of fundamental photon wavevectors, resulting in TH emission along the pump propagation axis.

The remaining two TH k-vectors at smaller angles, correspond to the non-collinear phase-matching configurations. A straightforward vectorial analysis based on photon momentum conservation enables the calculation of the exact TH emission angles ($\theta_{TH}$) using the following expression:

$$\theta_{TH}(\theta_F, n_l, n_r) = \tan^{-1}\left(\frac{n_l - n_r}{n_l + n_r}\tan(\theta_F)\right) \quad (1)$$

Here, $\theta_F$ denotes the angle of incidence of the fundamental beams, and $n_l$ and $n_r$ represent the number of photons "donated" from the left and right beams, respectively. Only combinations satisfying the condition $n_l + n_r = 3$, yield valid TH emission angles, consistent with energy conservation. Each configuration corresponds to a distinct phase-matching pathway, allowing a precise prediction of the TH emission direction via momentum conservation.

With regards to point ii), the combined effect of two-beam interference and pronounced nonlinearities in low-index media gives rise to a transient grating oriented perpendicular to the structural metagrating (see sample schematic in Fig. 1a). This optically induced periodic structure also diffracts the generated TH signal, as illustrated in Fig.2a-3 and Fig.2a-4, where the resulting diffraction orders are labeled by indices $m$ and $n$ for the inner and outer fundamental TH beams, respectively. As previously done for the emission angle pertaining photon momentum conservation, a simplified angular analysis can also be applied to the diffraction orders arising from the transient temporal grating. By applying the standard grating equation to generated TH light, we attain the following expression for the diffracted TH angle ($\alpha_{TH}$):

$$\alpha_{TH}(\theta_{TH}, n) = \sin^{-1}\left(\frac{n\lambda_{TH}}{\Lambda} - \sin(\theta_{TH})\right) \tag{2}$$

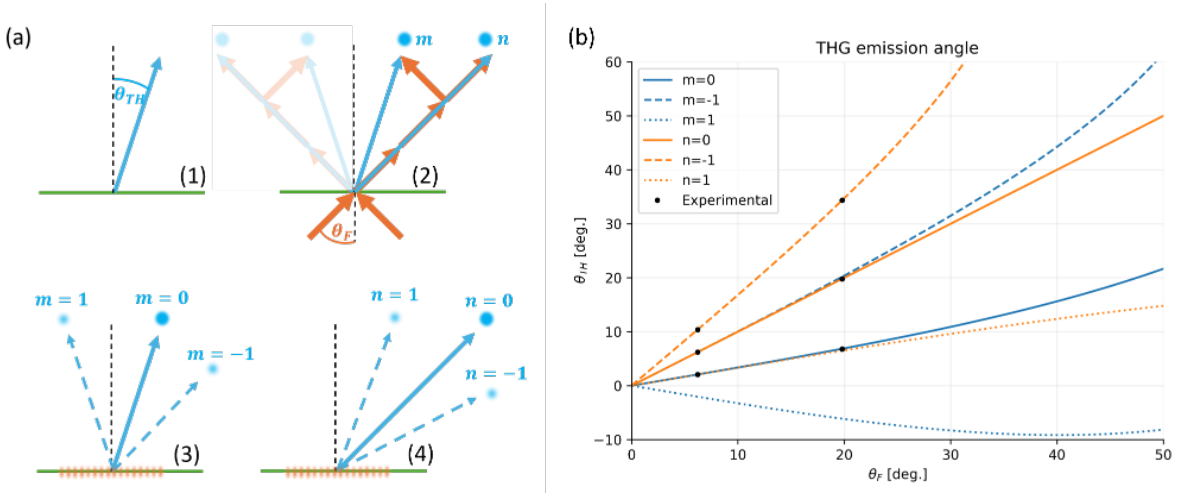

**Figure 2**: **Third harmonic angular analysis.** (a) Schematic of the mechanism determining the angular distribution of TH emission. Panels (1) and (2) show the phase matching geometry for fundamental TH beams produced by two-beam interaction inside the AZO films where $\theta_{TH}$ indicates the output angle between a given TH ray and the surface normal. Panels (3) and (4) illustrate how each phase matched TH beam is subsequently diffracted by the temporal grating; only the first diffraction orders are shown for clarity (indices n=±1 and m=±1). (b) Calculated TH emission angle $\theta_{TH}$ for all considered beams (n=0,±1 and m=0,±1) plotted as a function of the pump incidence angle $\theta_F$. The curves demonstrate that for pump incidence below 30°, several diffracted TH rays nearly coincide the phase-matched emission directions, producing strong overlap between diffraction and intrinsic THG channels.

Here, $n$ (or $m$) denotes the diffraction order, $\Lambda$ is the groove spacing, $\lambda_{TH}$ is the third-harmonic wavelength with $\lambda_{TH} = \lambda_F/3$ ($\lambda_F$ being the fundamental pump wavelength), and $\theta_{TH}$ is the input TH angle as defined by Eq. 1. The period of the transient grating produced by the interference of the two pump beams is therefore:

$$\Lambda = \frac{\lambda_F}{2\sin(\theta_F)} = \frac{3\lambda_{TH}}{2\sin(\theta_F)} \tag{3}$$

If now we focus solely on the collinear TH labeled $n$, for which $\theta_{TH} = \theta_F$, and substituting Eq. 3 into Eq. 2, we obtain the following formula linking the angle of the diffraction order with the incident angle:

$$\alpha_{TH}(\theta_F, n) = \sin^{-1}\left(\left(\frac{2n}{3} - 1\right)\sin(\theta_F)\right) \tag{4}$$

Evaluating $\theta_{TH}$ and $\alpha_{TH}$ from Eq. 1 and Eq. 4 for two experimental incidence angles below 30° (i.e., 6° and 20°) reveals near-perfect angular coincidence between the phase-matched and diffracted TH beams. Fig. 2b shows a complete view of the output angle $\theta_{TH}$ as a function of the input angle $\theta_F$ for all fundamental TH phase-matched beams and their first diffraction orders ($m = 0, \pm 1$ and $n = 0, \pm 1$).

We now turn to Fig. 1b-4, where the most revealing feature is the central horizontal band corresponding to the zero spatial-diffraction order, where six TH signals are recorded. The four central spots arise from a combination of nonlinear phase matching and temporal grating diffraction, as discussed above. The two outermost beams along the horizontal line, observed at the largest diffraction angles, cannot be accounted for by Eq. 1 or by higher-order wave-mixing; we, therefore, attribute their origin exclusively to temporal-grating diffraction. Notably, the temporal $n = -1$ diffraction order exhibits an intensity comparable to that of the spatially diffracted counterpart, indicating that the transient grating effectively modulates the observed TH emission, mirroring the influence of a physical grating.

When comparing Fig. 1b-3 and Fig. 1b-4, the absence of the outermost TH spots in the former is explained by the generation enhancement induced by the metagrating: those diffraction orders exceed the detection threshold in two-beam-on-grating configuration. This interpretation is supported by the experimental data in Fig. 3, which report measured TH for all four experimental configurations (i.e., single beam on unpatterned film, single beam on grating, two beams on unpatterned film, and two beams on grating). Here conversion efficiency is defined as the ratio of the forwardly emitted TH energy normalized to the incident fundamental energy (for the two-beam case the total incident energy of both pumps is used). The measurements show a pronounced efficiency increase when the grating is present, which we attribute to the strong local field enhancement from gap modes.

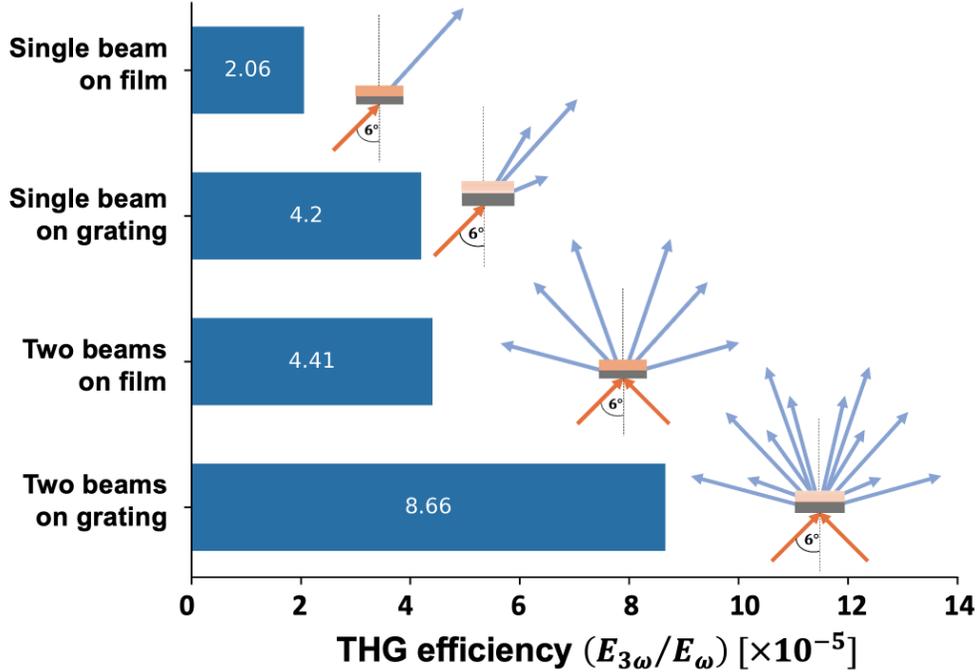

**Figure 3**: **Efficiencies of third harmonic generation in various configurations**. From the top down we have the total transmitted THG efficiency from a: single beam on unpatterned film; single beam on meta-grating; two beams on unpatterned film; two beams on meta-grating. Efficiency calculations include all diffracted THG orders and all incoming fundamental energy (in the two-beam configuration the total energy of both beams is considered so the input energy is twice the case of the single beam).

As both temporal grating line density and TH wavelength are proportional to the incident NIR wavelength, Eq. 4 becomes wavelength independent. We validate this experimentally by tuning the pump wavelength and recording the resulting diffraction pattern (Fig. 4a). The data confirm that the temporally diffracted beams remain fixed in angle as the wavelength changes, whereas beams diffracted by the static spatial grating shift vertically (dashed lines in Fig. 4a highlight this motion). Measured beam positions (blue dots) match the theoretical predictions (red circles computed from Eq. 1 and Eq. 4) with excellent agreement. Corresponding laboratory images (Fig. 4a-3 and Fig. 4a-4) are shown below the plots to provide a clear representation of the process while also giving a direct visual comparison between temporal and structural scattering.

Thus far we have explored the spatio-temporal metasurface by varying wavelength and incidence angle. Because the system is intrinsically time-dependent, time itself can be used as one of the control parameters. To demonstrate this, we measured the device response as a function of the delay between the two trains of pump pulses. The colormap in Fig. 4b-1 shows spectral measurements of the generated TH signal labeled "$m$" in Fig. 2, recorded versus pump–pump delay at an incidence angle of 6°. From these data we tune the TH wavelength by few nanometers, as highlighted in Fig. 4b-2, which displays two representative spectra from Fig. 4b-1 at delays $\tau = 67$ and $\tau = 0$ fs. Although the delay-controlled tuning observed here is modest, it stems from a time-refraction mechanism that can produce substantially larger shifts, up to 40 nm under optimized conditions (*38*).

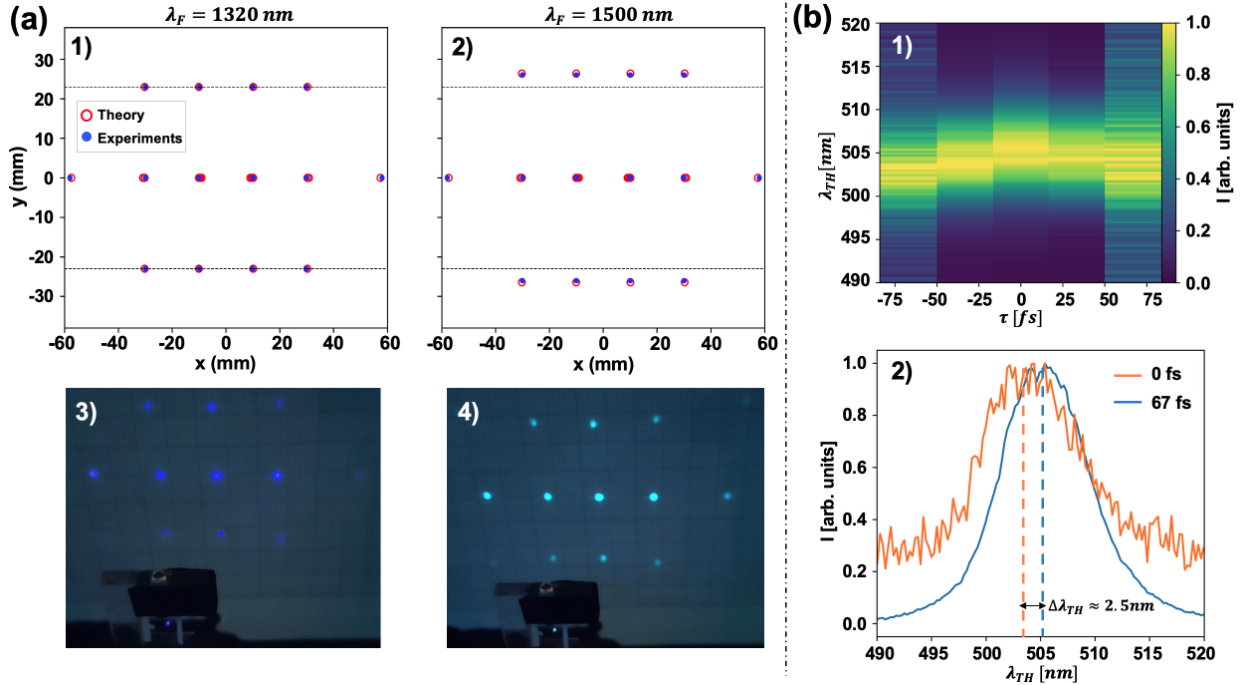

**Figure 4**: **Spatial-temporal metagrating tuneability via wavelength and pumps delay.** (a) Experimental (blue dots) and theoretical (red circles) positions of TH beams at the detection screen at 1320 nm (panel 1) and 1500 nm (panel 2) for a 20° incidence angle. The figure also reports correspondent real life laboratory pictures of emitted TH signal (panel 3 and panel 4). The magnitude of scattering from both spatial and temporal structures are comparable. (b) In panel 1 we report experimentally acquired TH intensity spectra as a function of several time-delays between the two pump beams at 1500 nm. The temporal delay between the two interfering trains of incident pulses is demonstrated to be an additional design parameter for the temporal lattice. Panel 2 shows intensity spectra for two specific time-delay to underline the TH tunability range.

## Discussion

In this work we perform a direct comparison between scattering from permanently fabricated nanostructures and analogous nanopatterns imprinted transiently by optical interference on time-varying films. We selected a grating geometry because of its fundamental role and broad versatility in optics and photonics. To isolate scattering contributions arising from different origins, the permanent structural grating and the optically induced transient gratings are oriented orthogonally, so that their diffraction signatures separate clearly in the far field. Moreover, by leveraging large low-index nonlinearities, we do not need to use an external probe since the system self-produces a short-wavelength probe via THG which resolves the nanoscale features of our spatio-temporal metasurface. This yields a complete x-y diffraction map at the detector, in which structural and temporal diffraction orders appear with comparable intensities.

The grating device also serves to remarkably enhance THG compared with flat films via local field enhancement. Beyond the static, fabricated pattern, the non-stationary nature of our metasurface opens a multi-dimensional design space: the spatio-temporal meta-lattice does not need any lithographic step but is dynamically modulated by means of the pump wavelength, angles of incidence, and temporal delay between pulse trains. Tuning these knobs reshapes the effective 2D-meta-cell while also tuning the self-generated probe signal, producing large changes in amplitude, phase, spectrum, and momentum of the scattered light.

Our experiments exploit highly nonlinear, low-index conductive oxides to realize metagrating with exceptional ultra-fast reconfigurability. Unlike purely index-modulation studies, we focus on optically induced scattering, a direct signature of light matter interaction that permits quantitative benchmarking of transparent conductors for temporal lithography. This approach moves device design from fixed fabrication constraints toward radiation engineering, adding real-time tunability on femtosecond timescales suggesting new routes to ultrafast reconfigurable metasurfaces, adaptive spectral and spatial multiplexers, and engineered high-harmonic sources.

# Acknowledgments

All authors would like to thank Dr. Aaron Naden from the School of Chemistry at the University of St Andrews for support with the fabrication of the samples with the FIB.